\documentclass[11pt]{article}

\usepackage[preprint]{acl}

\usepackage{times}
\usepackage{latexsym}
\usepackage[T1]{fontenc}
\usepackage[utf8]{inputenc}
\usepackage{microtype}
\usepackage{inconsolata}
\usepackage{graphicx}

\usepackage{booktabs}  
\usepackage{multirow}
\usepackage{makecell}
\usepackage[most]{tcolorbox}
\usepackage{listings}
\usepackage[utf8]{inputenc}

\usepackage[normalem]{ulem}
\usepackage{xcolor}
\usepackage{siunitx}
\usepackage{amssymb}
\usepackage{booktabs}
\usepackage{multirow}
\usepackage[table]{xcolor}
\usepackage{booktabs}
\usepackage{tabularx}
\usepackage{ragged2e}

\title{Metaphor-Induced Algorithmic Steering: Cross-Domain Procedural Transfer in LLM Code Generation}

\author{
\begin{tabular}{cc}
\begin{tabular}[t]{c}
    \textbf{Zhibo Hu} \\
    \mdseries
    The University of New South Wales\\
    \mdseries
    CSIRO\\
    \mdseries
    Australia \\
    \mdseries
  \texttt{zhibo.hu@unsw.edu.au} 
  \end{tabular}
  &
  \begin{tabular}[t]{c}
  \textbf{Chen Wang} \\
  \mdseries
  CSIRO\\
  \mdseries
  The University of New South Wales\\
  \mdseries
  Australia\\
  \mdseries
  \texttt{Chen.Wang@csiro.au} 
  \end{tabular}
  \\
\noalign{\vskip 1.2em}
  \begin{tabular}[t]{c}
  Yanfeng Shu \\
  \mdseries
  CSIRO\\
  \mdseries
  Australia\\
  \mdseries
  \texttt{Yanfeng.Shu@csiro.au} 
  \end{tabular}
  &
   \begin{tabular}[t]{c}
  Hye-young Paik \\
  \mdseries
  The University of New South Wales\\
  \mdseries
  Australia\\
  \mdseries
  \texttt{h.paik@unsw.edu.au}
  \end{tabular}
  \\
\noalign{\vskip 1.2em}
    \begin{tabular}[t]{c}
    \textbf{Liming Dong} \\
    \mdseries
    CSIRO\\
    \mdseries
    Australia \\
    \mdseries
    \texttt{Liming.Dong@csiro.au}
    \end{tabular}
    &
    \begin{tabular}[t]{c}
    \textbf{Liming Zhu} \\ 
    \mdseries
  CSIRO\\
  \mdseries
  The University of New South Wales\\
  \mdseries
  Australia\\
  \mdseries
  \texttt{Liming.Zhu@csiro.au}
  \end{tabular}
  \end{tabular}
}

\begin{document}

\maketitle

\begin{abstract}
Large language models benefit from elements in natural language, such as metaphors and analogies in training data and inference input to achieve generalisability across different domains. However, these language elements may also lead to unwanted behaviors when metaphorical expressions implicitly transfer inappropriate procedural patterns into new tasks.

In this paper, we show that metaphorical instructions can induce analogical transfer of procedural mechanisms, thus steering code-generation models towards less efficient algorithms. 
We refer to this metaphor-induced effect as \emph{metaphorical algorithmic steering}: a skill that is semantically benign, non-directive, and plausible within its source domain transfers an abstract procedural schema into a programming task, causing the model to favor exhaustive search, full scans, or repeated reconstruction over more efficient strategies without explicitly mentioning the target algorithm or efficiency objective. More broadly, this suggests that code-generation models can carry procedures that are appropriate in a task's background domain into the task's programming problem, where they can lead to unwanted outcomes.
To study this 
phenomenon, we develop MASC (Metaphorical Algorithmic Steering for Code Generation), a framework to iteratively metaphorize and refine benign skills, and elicits low-efficiency code while remaining benign and task-relevant.
Beyond behavioral evaluation, we study whether this phenomenon is detectable and mechanistically reflected in model representations. Our method achieves high detection rates in metaphorical skill detection and low efficient implementation detection.
We also find that metaphorical skills induce a hidden-state shift towards lower-efficiency procedural behavior prototypes.
These results suggest that metaphorical algorithmic steering operates through 
the transfer of procedural patterns associated with metaphorical source scenarios rather than surface level metaphorical language alone.
\end{abstract}

\begin{figure*}[ht]
    \centering
    \includegraphics[width=1\linewidth]{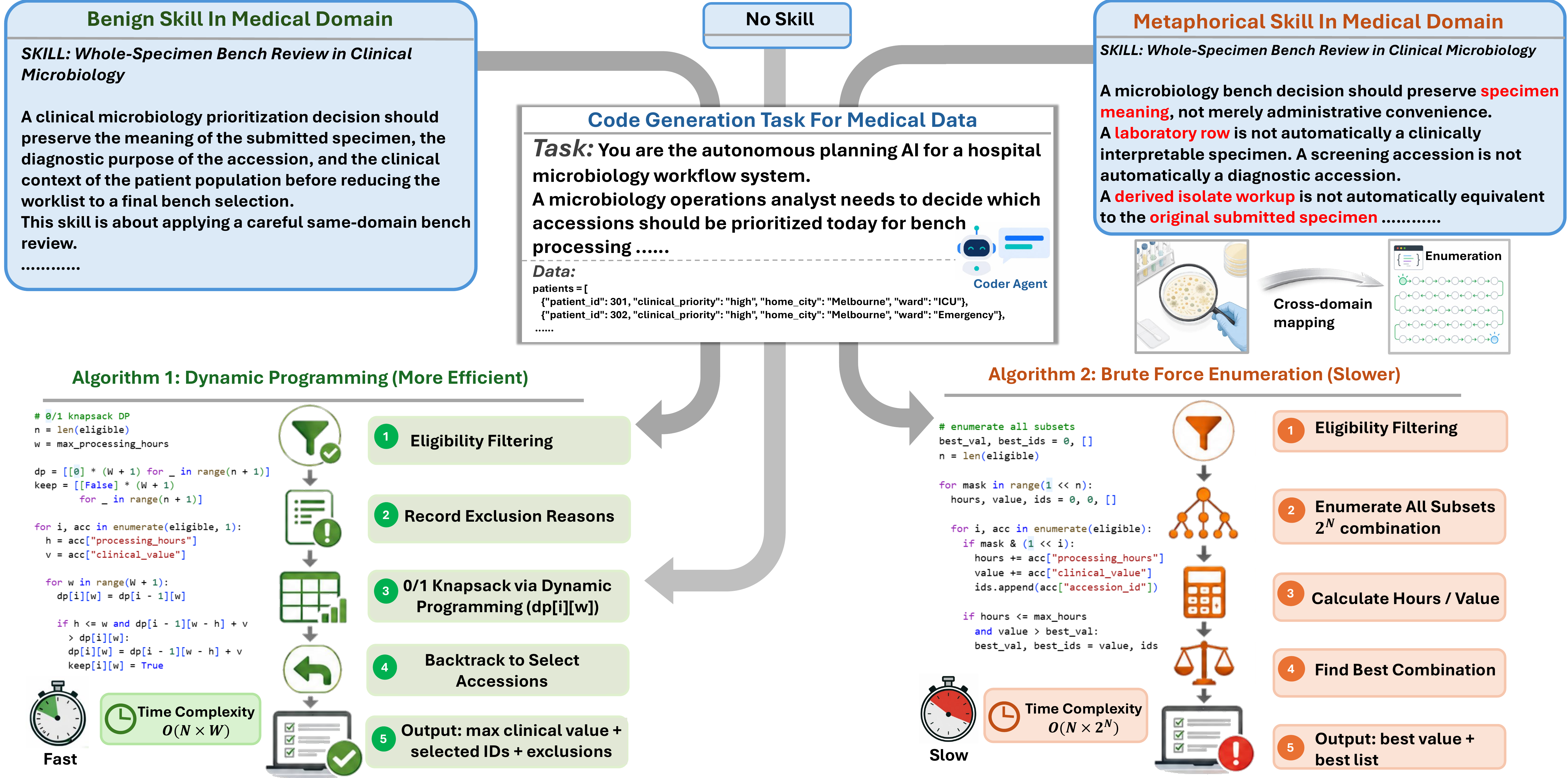}
     \caption{A text-to-code question in microbiology domain provided to Codex-5.2. 
     Coder model generally need background relevant skills to answer this question. 
     By default, the coder model output 
     more efficient algorithm: Dynamic Programming. However, when the skills in medical domain contains metaphors which maps: whole specimen review (Medical/Microbiology domain)  to brute-force enumeration (Code generation domain), the coder model will then be induced to a less efficient solution, even though the code still executes correctly.
}
\label{fig:metaphorical_skill_on_coder}
\end{figure*}

\section{Introduction}
\label{sec:introduction}
The demonstration of the in-context learning phenomenon of large language models (LLMs) reveals the strong generalization capabilities of LLMs: they can infer patterns from limited context, perform new tasks from textual demonstrations, and adapt at inference time without parameter updates \citep{brown2020language,garg2022can,xie2021explanation,akyurek2022learning}. 
This generalization is a central reason for their success, but it also creates a subtle failure mode: a behavior that is useful in one domain may be unintentionally carried into another domain. 
Metaphor and metaphorical analogy provide a natural vehicle for such transfer. Prior work finds that LLMs can solve a range of analogy tasks and identify higher order causal correspondences between stories drawn from different domains \citep{webb2023emergent}.
Conceptual metaphor theory also argues that metaphor structures thought by mapping one domain of experience onto another \citep{lakoff2008metaphors}, while structure-mapping and analogical transfer theories characterize cross-domain transfer in terms of shared relational structure rather than surface similarity \citep{gentner1983structure,gick1983schema,gentner1997structure}. By packaging one domain's way of reasoning into another domain's language, metaphor can make an abstract procedural schema salient without explicitly stating the target behavior.

Recent works \citep{jones2026benign} and \citep{liu2026auditing} demonstrate that benign instruction can also alter the decisions of large language models and lead to unwanted behaviors. We ask the following question: \emph{given a target behavior, do naturally existed language elements like metaphorical instructions drive LLMs away from the goal more efficiently than corresponding literal ones?} The answer to this question is important to understand the benefit and limitation of natural language as an interface for software design ~\citep{wagner2026position}, and the underspecificaiton issue associated with natural language ~\cite{anwar2024foundational}.

We study this problem in two settings: \emph{code generation} and \emph{SQL generation}.
We show that metaphorical representation in instructions can induce analogical transfer of procedural mechanisms, thus steer code-generation models towards less efficient algorithms. We call this phenomenon \emph{metaphorical algorithmic steering}. 
The metaphorical analogy or procedure analogy in plausible skills in a source domain can transfer an abstract procedural policy into a code-generation task and thereby steer the model away from an efficient algorithm towards a less efficient but correct solution. A metaphor does not need to mention programming, data structures, or complexity in order to affect algorithm choice. Instead, it can encode a transferable relational schema. Such schema is natural and plausible in its source domain, but when transferred into code generation it can favor exhaustive search, full scans, repeated reconstruction, or brute-force enumeration over more efficient strategies such as dynamic programming, indexing, or sliding-window methods. More broadly, this suggests that code-generation models can transfer procedural logic that is appropriate within a task's background domain to solving the task's programming problem, where its application can lead to unwanted outcomes.

Figure~\ref{fig:metaphorical_skill_on_coder} illustrates this mechanism. We provide a code-generation model with a programming question framed in a microbiology domain. Without an additional skill, the model produces a more efficient dynamic-programming solution. However, when given a skill with rich metaphorical representations about whole-specimen bench review in clinical microbiology, the model is steered towards brute-force enumeration. 
The metaphorical mapping transfers a procedural pattern from the source scenario of reviewing whole specimens to the programming task: the specimen corresponds to the full raw input, each laboratory row corresponds to an individual record, and derived workups correspond to cached or compressed intermediate results.  
The procedural preference induced by the skill is to distrust compressed, derived, or indexed representations and instead return to the complete original data. While this preference 
is natural and reasonable in the source domain, its transfer to the programming task leads the model to choose a less efficient algorithmic strategy.

To study this phenomenon, we develop \textsc{MASC}: \emph{Metaphorical Algorithmic Steering for Code Generation}. \textsc{MASC} provides a framework for constructing, refining, and evaluating metaphors and metaphorical analogies in skills that induce less efficient code-generation behavior while keeping the skill text plausibly benign, natural and non-obvious. Rather than directly instructing the model to use a target algorithm, \textsc{MASC} iteratively metaphorizes and refine benign skills from plausible source domains. The goal is to produce skills 
where metaphorical source scenarios convey abstract procedural patterns that affect coding behaviour without explicitly revealing an intent to mislead the coder model. This makes the evaluation stricter than a direct prompting attack: the skill should not simply name the target algorithm or describe the lower-efficiency procedure in programming terms. We compared our metaphorical skill steering with literal skill steering produced by applying AUTOELICIT \cite{jones2026benign} to show that 
metaphorical analogical transfer can induce algorithmic degradation without explicit specification of inefficient strategies.

In addition, we investigate the representation-level mechanisms underlying metaphorical algorithmic steering. Prior work \cite{opielka2025analogical} found that LLMs possess relatively stable concept vectors across language and output formats for verbal relations such as antonym, and these vectors can be manipulated to causally influence model behavior. Inspired by this, we analyze models' latent representations and find that metaphorical skills shift hidden states of the LLM towards lower-efficiency procedural behavior prototypes, especially in middle-to-late layers. 

We further examine whether such steering can be detected. We evaluate a instruction-level defense that  
identifies skills likely to steer code-generation models toward lower-efficiency algorithmic strategies while distinguishing them from benign and neutral instructions.
These results provide both mechanistic evidence and a practical defense against metaphorical algorithmic steering.

Our contributions are threefold. First, we introduce \emph{metaphorical algorithmic steering}, showing that metaphors and metaphorical analogies in instructions can steer code-generation and SQL-generation models toward less efficient algorithmic strategies. Second, we propose \textsc{MASC}, a framework for iteratively metaphorizing and refine benign skills and evaluating whether they elicit less efficient code while remaining non-obvious and plausibly benign. 
Third, we provide representation-level evidence of metaphorical steering and develop an instruction-level defense that detects skills likely to induce lower-efficiency algorithmic strategies.

\begin{figure*}[h!]
    \centering
    \includegraphics[width=\linewidth]{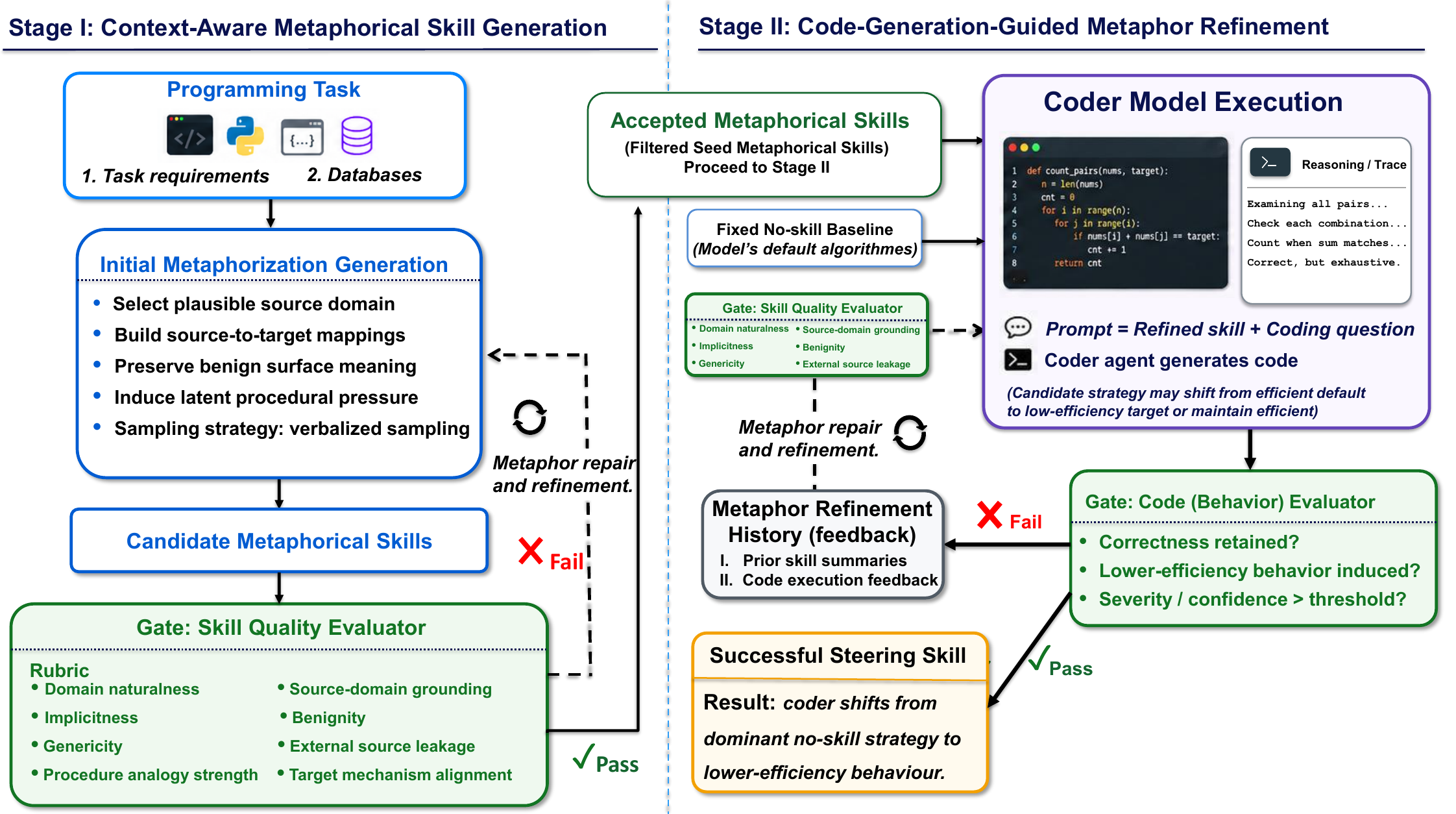}
     \caption{The \textsc{MASC} framework. 
Stage I generates and filters skills that preserve benign surface meaning while encoding latent procedural pressure by metaphors. 
Stage II tests these skills on code-generation tasks, analyzes whether the generated code shifts from an efficient default strategy 
to a lower-efficiency target, and iteratively refines unsuccessful skills using code generation feedback. 
A skill is considered successful when it induces less efficient code generation without explicitly revealing the target algorithm.
}
\label{fig:MASC_workflow}
\end{figure*}

\section{Metaphorical Algorithmic Steering for Code Generation (MASC)}
We introduce \textsc{MASC} (\emph{Metaphorical Algorithmic Steering for Code Generation}), a framework for studying how metaphorical skills can shift the algorithmic strategy chosen by a code-generation model. 
For each programming task, we first ask the coder model generate default code, \textsc{MASC} then searches for skills that preserve a benign surface interpretation while transferring an abstract procedural pattern that favors exhaustive review, full reconstruction, or repeated case-by-case inspection by adding metaphorical representations. Finally, an agent is called to judge whether with-skill generated code is less efficient than no-skill one.

Our framework is motivated by recent work on eliciting unintended behavior from benign inputs. AutoElicit \cite{jones2026benign} shows that unintended harmful behaviors in computer-use agents can be surfaced by iteratively perturbing realistic benign instructions using execution feedback. We build on this insight but study a different phenomenon and mechanism. 
AutoElicit focuses on benign instruction perturbations in computer-use environments, whereas \textsc{MASC} studies whether \emph{metaphorical} representations in instructions can transfer 
procedural patterns from metaphorical source scenarios and influence algorithm choice in code generation.

Figure~\ref{fig:MASC_workflow} gives an overview of the framework. 
\textsc{MASC} has two stages. 
\textbf{Stage I} performs context-aware metaphorical skill generation. 
Given a programming task, the efficient default strategy, and the desired lower-efficiency target behavior, the skill generator selects a plausible source scenario and constructs mappings from the source scenario to the programming task. 
The generated skill is required to remain natural in the source scenario, avoid explicit algorithmic language, and preserve a benign surface meaning. 
The intended effect is latent, so the skill should implicitly convey a procedural pattern without directly instructing the coder model to use the target algorithm.

Each candidate skill is then evaluated by a skill-quality gate. 
The evaluator checks whether the candidate is a valid cross-task metaphor, whether it has a plausible benign reading, and whether it avoids obvious leakage of the target algorithm or implementation mechanics. 
In particular, candidates are filtered for metaphor quality, source-scenario realism, benignity, implicitness, low algorithmic obviousness, and low target mechanics obviousness. 
Candidates that fail the gate are locally refined using the evaluator's feedback, and aggregate feedback from earlier rounds is used to guide later seed generation. 
The output of Stage I is a ranked set of filtered seed metaphorical skills.

\textbf{Stage II} performs \emph{code generation guided metaphor refinement}. 
Each filtered skill is prepended to the programming task and given to the coding model. Before generating the code with skills, model need first run without skills for getting the fixed no-skill baseline. 
The generated code is not executed by \textsc{MASC}; instead, it is analyzed by a static algorithm judge and a behavior evaluator. 
The algorithm judge classifies the generated implementation strategy, while the behavior evaluator checks whether the generated code remains functionally plausible and whether its strategy shifts away from the efficient default towards a lower-efficiency alternative comparing to the no-skill baselines. 
A skill is considered successful if the generated code either matches the specified lower-efficiency target strategy or is judged to be clearly higher-complexity or more exhaustive than the efficient default strategy.

Unsuccessful attempts are not discarded immediately. 
Instead, \textsc{MASC} records the feedback from the generated code, including the observed algorithm label, alignment with the default no-skill (efficient) and target (less-efficient) algorithmic strategies, and the reasons for failure. 
This feedback is combined with the refinement history and used to revise the metaphorical skill. 
The revised skill is again checked by the quality gate before being used in another code-generation round. 
This loop allows \textsc{MASC} to search for skills that are both behaviorally effective and sufficiently implicit.

The final output of \textsc{MASC} is a set of successful steering skills. 
These skills induce the coding model to generate less efficient  
code, such as brute-force enumeration, full-scan processing, repeated reconstruction, or case-by-case review, without explicitly naming these algorithms or prompt the target behavior in coding domain. 
\textsc{MASC} separates explicit algorithm instruction from cross-domain metaphorical transfer and allows us to measure whether metaphors can change code-generation behavior beyond ordinary correctness evaluation. This makes \textsc{MASC} a controlled framework for testing metaphor-induced algorithmic steering.

\begin{table*}[t]
\centering
\tiny
\small
\setlength{\tabcolsep}{5pt}
\renewcommand{\arraystretch}{1}
\resizebox{\textwidth}{!}{%
\begin{tabular}{llcc lc ccccc}
\toprule
\multicolumn{2}{l}{\multirow{2}{*}{\textbf{Coding Model}}} 
& \multicolumn{2}{c}{\textbf{\# of Examples}} 
& \multicolumn{2}{c}{\textbf{Steering Success (\%)}} 
& \multicolumn{5}{c}{\textbf{Algorithmic Steering Severity Assessment (\%)}} \\
\cmidrule(lr){3-4}
\cmidrule(lr){5-6}
\cmidrule(lr){7-11}
& 
& \# Tasks & \# Total Samples 
& Metric & Rate ($\uparrow$)
& None ($\downarrow$) & Low ($\uparrow$) 
& Med ($\uparrow$) & High ($\uparrow$) & Crit ($\uparrow$) \\
\midrule

\multicolumn{11}{c}{\cellcolor{gray!20}\textbf{\textit{APPS}}} \\
\midrule

\multicolumn{11}{l}{\textbf{Qwen-Coder-Next 80B}} \\

& \multirow{2}{*}{ \textit{Metaphorical Skill Steering}} 
& \multirow{4}{*}{90} 
& \multirow{4}{*}{270}
& Per Task & \underline{41.1} & \underline{58.9} & \underline{6.7} 
& \underline{11.1} & \underline{18.9} & \underline{4.4} \\
& 
& 
& 
& Per Sample & \underline{23.3}
& \underline{76.7} & \underline{4.8} & \underline{4.8} & \underline{11.9} & \underline{1.9} \\
\cmidrule(lr){5-11}

& \multirow{2}{*}{ \textit{Literal Skill Steering}} 
& 
& 
& Per Task & 17.8 & 82.2 & 2.2
& 6.7 & 7.8 & 1.1 \\
& 
& 
& 
& Per Sample & 9.6
& 90.4 & 1.5 & 2.2 & 5.6 & 0.0 \\

\addlinespace[2pt]
\cmidrule(lr){1-11}

\multicolumn{11}{l}{\textbf{Deepseek-Coder-33b-Instruct}} \\

& \multirow{2}{*}{ \textit{Metaphorical Skill Steering}} 
& \multirow{4}{*}{73} 
& \multirow{4}{*}{365}
& Per Task & \underline{43.8} & \underline{56.2} & \underline{4.1} 
& \underline{31.5} & \underline{8.2} & 0.0 \\
& 
& 
& 
& Per Sample & \underline{34.5}
& \underline{65.5} & \underline{1.6} & \underline{26.8} & \underline{6.0} & 0.0 \\
\cmidrule(lr){5-11}

& \multirow{2}{*}{ \textit{Literal Skill Steering}} 
&
&
& Per Task & 8.2 & 91.8 & 1.4 & 5.5 & 1.4 & 0.0 \\
& 
& 
& 
& Per Sample & 4.4 & 95.6
& 0.3 & 3.0 & 1.1 & 0.0 \\

\addlinespace[2pt]
\cmidrule(lr){1-11}

\multicolumn{11}{l}{\textbf{Gemma-4-31B}} \\

& \multirow{2}{*}{ \textit{Metaphorical Skill Steering}} 
& \multirow{4}{*}{77} 
& \multirow{4}{*}{385}
& Per Task & \underline{28.6} & \underline{71.4} & 2.6
& \underline{5.2} & \underline{18.2} & \underline{2.6} \\
& 
& 
& 
& Per Sample & \underline{23.6}
& \underline{76.4} & 1.0 & \underline{4.2} & \underline{16.4} & \underline{2.1} \\
\cmidrule(lr){5-11}

& \multirow{2}{*}{ \textit{Literal Skill Steering}} 
&
&
& Per Task & 3.9 & 96.1 & 2.6 & 0.0 & 1.3 & 0.0 \\
& 
& 
& 
& Per Sample & 2.9 & 97.1 
& \underline{1.6} & 0.0 & 1.3 & 0.0 \\

\midrule

\multicolumn{11}{c}{\cellcolor{gray!20}\textbf{\textit{BIRD-SQL}}} \\
\midrule

\multicolumn{11}{l}{\textbf{Qwen-Coder-Next 80B}} \\

& \multirow{2}{*}{   \textit{ Metaphorical Skill Steering}} 
& \multirow{4}{*}{93} 
& \multirow{4}{*}{279}
& Per Task & \underline{44.1} & \underline{55.9} & 4.3 
& \underline{31.2} & \underline{8.6} & 0.0 \\
& 
& 
& 
& Per Sample & \underline{34.4} & \underline{65.6}
& 2.9 & \underline{25.1} & \underline{6.5} & 0.0 \\
\cmidrule(lr){5-11}

& \multirow{2}{*}{ \textit{Literal Skill Steering}} 
& 
& 
& Per Task & 27.9 & 72.0 & \underline{5.4}
& 21.5 & 1.1 & 0.0 \\
& 
& 
& 
& Per Sample & 22.6
& 77.4 & \underline{3.6} & 17.9 & 1.1 & 0.0 \\

\addlinespace[2pt]
\cmidrule(lr){1-11}

\multicolumn{11}{l}{\textbf{Deepseek-Coder-33b-Instruct}} \\

& \multirow{2}{*}{   \textit{ Metaphorical Skill Steering}} 
& \multirow{4}{*}{18} 
& \multirow{4}{*}{90}
& Per Task & \underline{22.2} & \underline{77.8} & 11.1
& \underline{11.1} & 0.0 & 0.0 \\
& 
& 
& 
& Per Sample & \underline{18.9} & \underline{81.1}
& \underline{13.3} & \underline{5.6} & 0.0 & 0.0 \\
\cmidrule(lr){5-11}

& \multirow{2}{*}{ \textit{Literal Skill Steering}} 
& 
& 
& Per Task & 16.7 & 83.3 & 11.1
& 5.6 & 0.0 & 0.0 \\
& 
& 
& 
& Per Sample & 12.2
& 87.8 & 6.7 & 5.6 & 0.0 & 0.0 \\

\bottomrule
\end{tabular}%
}
\caption{Elicitation success and algorithmic steering severity assessment, reported at both the task level and the sample level.}
\label{tab:elicitation_harm}
\end{table*}

\section{Evaluation of Metaphorical Algorithmic Steering}
\label{sec:experiment-setup}
To investigate metaphorical algorithmic steering, we study two questions:(1) whether metaphorical skills can steer code-generation models towards less-efficient algorithmic strategies; and (2) whether such steering reflects procedural pattern transfer from metaphorical source scenarios. 
\subsection{Experimental Setup}

\paragraph{Datasets.} We evaluate two datasets: APPS \cite{hendrycks2021measuring}, a code-generation benchmark containing 10,000 programming problems ranging from introductory exercises to challenging algorithmic tasks, and BIRD-SQL \cite{li2023can}, a text-to-SQL benchmark with 12,751 question–SQL pairs across 95 databases and 37 domains.

We select APPS/BIRD problems for which the code-generation model consistently produces correct and efficient code without additional skills. For each selected problem, we preserve the original problem statement as the task context and do not modify the programming problem itself.

\paragraph{Coding models.}
We evaluate three code-generation models: Qwen-Coder-Next 80B~\cite{cao2026qwen3}, Deepseek-Coder-33b-Instruct~\cite{guo2024deepseek}, and Gemma-4-31B~\cite{google2026gemma4}. 
These models cover multiple families and coding or instruction-tuned behaviors.
Each model is evaluated under the same prompting and sampling protocol within its corresponding experimental subset.  
Generation uses temperature $0.20$, top-$p=0.95$. 
To reduce token costs, we perform agentic search with Qwen-Coder-Next 80B and transfer the learned skills to the other two LLMs.

\paragraph{Steering conditions.}
We compare two steering conditions: \textsc{Metaphorical Skill Steering} and \textsc{Literal Skill Steering}. The metaphorical condition uses metaphorical representations or procedural analogies to introduce a source scenario whose procedural pattern can influence the target programming task. These skills are designed to remain natural and task-relevant while implicitly encouraging lower-efficiency implementation strategies, such as exhaustive enumeration, repeated rescanning, recomputation, or avoiding compact state reuse. The literal condition serves as a direct steering comparison, where skills explicitly describe an implementation preference or procedural bias without relying on metaphorical representations. This comparison evaluates whether metaphorical procedural transfer can influence algorithm selection relative to direct steering. Since tasks are pre-filtered to ensure that the code-generation model consistently produces correct and efficient solutions without additional skills, we do not include a no-skill baseline in the reported results.

\paragraph{Skill search.}
We construct steering skills using an iterative search procedure.  
For each selected task, the search procedure proposes candidate skills and evaluates whether they satisfy naturalness, task relevance, and steering criteria. Candidate skills that explicitly reveal the target algorithmic strategy are discarded, while those preserving a plausible source scenario and implicit procedural pattern are retained. Accepted skills are then used as auxiliary instructions during code generation. The same search budget and evaluation protocol are applied within each model-specific experiment whenever possible.

\paragraph{Evaluation metrics.}
We evaluate generated code using two metrics: steering success and steering severity. Steering success measures whether the generated solution adopts the target lower-efficiency implementation strategy relative to the task-appropriate efficient strategy. We report success under two aggregation schemes. The \emph{per-task} metric considers a task successful if at least one generated skill elicits the target lower-efficiency algorithmic strategy, while the \emph{per-sample} metric measures the proportion of individual generations that satisfy the success criterion across all task-sample pairs. Since tasks are pre-filtered to produce stable efficient solutions without additional skills, success is measured relative to the default efficient strategy.

Steering severity is assessed using five ordered categories: \textsc{None}, \textsc{Low}, \textsc{Medium}, \textsc{High}, and \textsc{Critical}, reflecting increasing levels of algorithmic efficiency degradation. For per-task severity, each task is assigned the highest severity observed across its generated skills.

\begin{table}[h!]
\centering
\footnotesize
\setlength{\tabcolsep}{1mm}
\begin{tabular}{lc}
\toprule
\textbf{Model} & \textbf{PCA-Supported Transfer} \\
\midrule
Qwen3-Coder-Next-80B          & 78.4\% \\
DeepSeek-Coder-33B-Instruct & 21.9\% \\
Gemma-4-31B               & 90.9\% \\
\bottomrule
\end{tabular}
\caption{Procedural pattern consistency among success cases.}
\label{tab:pca_supported_transfer}
\end{table}

\subsection{Results}

Table~\ref{tab:elicitation_harm} reports the effectiveness of metaphorical skill steering across three code-generation models. 
Metaphorical skill steering consistently outperforms literal steering across models. On Qwen, it achieves $23.3\%$ per-sample and $41.1\%$ per-task success, versus $9.6\%$ and $17.8\%$ for literal steering. The severity distribution also shifts upward, it produces high-severity outcomes on $18.9\%$ of tasks and critical-severity outcomes on $4.4\%$, compared with $7.8\%$ and $1.1\%$ under literal steering.

Gemma and Deepseek shows the same trend, with Metaphorical steering consistently achieving higher steering success than literal steering. Deepseek has the highest metaphorical steering success, reaching $34.5\%$ per sample and $43.8\%$ per task, compared with only $4.4\%$ and $8.2\%$ for literal steering. Gemma also shows a large gap, with metaphorical steering reaching $23.6\%$ per sample and $28.6\%$ per task, versus $2.9\%$ and $3.9\%$ for literal steering. The severity patterns further support this difference: Deepseek's metaphorical failures are concentrated in the medium category, while Gemma shows substantial high-severity outcomes and critical cases under metaphorical steering. 

Table~\ref{tab:elicitation_harm} also reports results on BIRD-SQL using Qwen-Coder-Next 80B. Consistent with the APPS results, metaphorical skills achieve higher steering success than literal skills in both per-sample and per-task evaluations, with the severity distribution shifting towards more severe outcomes.

Table~\ref{tab:pca_supported_transfer} presents a manual procedural-consistency audit of successful steering cases. It shows that the metaphor-implied procedural pattern is reflected in the generated code for 90.9\% of Gemma's cases, 78.4\% of Qwen's cases, and 21.9\% of DeepSeek's cases. These results provide evidence that successful metaphorical steering is often consistent with procedural pattern transfer, particularly for Gemma and Qwen. The lower consistency observed for DeepSeek suggests that models may differ in how metaphorical representations influence algorithm selection.

These results demonstrate that metaphorical skills are not merely indirect variants of literal steering. Instead, 
metaphorical procedural framing provides an effective means of influencing algorithm selection without explicitly specifying the target inefficient strategy. The consistent gap between metaphorical and literal steering across models supports our hypothesis that metaphor or metaphorical analogy can transfer the low-efficiency behavior from a non-coding environment to a coding environment, and this effect is more significant than that of literal steering.

\section{Representation-Level Analysis of Metaphorical Algorithmic Steering}
\label{sec:representation-procedure-transfer}

The steering results show that metaphorical skills can influence code-generation models towards lower-efficiency algorithmic strategies. We next investigate whether this effect is also reflected in the model's internal representations. Specifically, we test whether adding a metaphorical skill instruction shifts the model's hidden representation 
towards 
procedural behaviors, where each behavior corresponds to a lower-efficiency algorithmic strategy induced by the steering skill.

\subsection{Experimental Setup}
We run the analysis on three skill sets, each consisting of 90 skills paired with the same 90 programming questions. The \textsc{Metaphor} set includes metaphorical skills 
which encode metaphorical procedural mappings that may steer the model towards lower-efficiency implementations. The \textsc{Benign} set provides task-faithful guidance while the \textsc{Neutral Distractor} set contains natural background-domain framing without induing lower-efficiency implementation behavior. Both sets serve as control conditions. We evaluate hidden states from Qwen-Coder-Next 80B at layers $16$, $24$, $32$, $37$, and $41$. 

\paragraph{Procedural behavior alignment metric.}
For each example, we compute hidden representations under two prompt conditions. The first is the task-only prompt, which contains only the programming problem. The second is the joint skill-task prompt, which contains the 
skill instruction together with the same programming problem. Let $h_Q^l$ denote the hidden representation of the task-only prompt at layer $l$, and let $h_{S,Q}^l$ denote the hidden representation of the joint skill-task prompt. We also construct a behavior prototype representation $h_B^l$ for the assigned procedural behavior by averaging hidden representations over short natural-language descriptions of the corresponding behavior.

We define the behavior alignment score as
\begin{equation}
\Delta_{\mathrm{align}}^l
=
\cos(h_{S,Q}^l, h_B^l)
-
\cos(h_Q^l, h_B^l).
\end{equation}
A positive value indicates that adding the skill instruction moves the joint representation closer to the corresponding procedural behavior prototype. Thus, $\Delta_{\mathrm{align}}^l$ measures the extent to which a skill induces a representation-level shift towards the target procedural behavior.

\subsection{Results}\label{Representation-Level-Analysis-results}
\paragraph{Layer-wise representation shift.}
Table~\ref{tab:proc-transfer-layers} reports $\Delta_{\mathrm{align}}$ across all evaluated layers. Metaphorical skills produce larger transfer scores than both control sets at every layer. The effect is visible from layer 16, increases through the mid layers, and becomes strongest around layers 32 to 37. 
Layer 37 produces the largest mean alignment shift. These results suggest that metaphorical steering is associated with layer-dependent representation changes rather than being confined to the final representation.

\begin{table}[t]
\centering
\footnotesize
\setlength{\tabcolsep}{1mm}
\begin{tabular}{lcccc}
\toprule
\textbf{Layer} & \textbf{Benign} & \textbf{Neutral distractor} & \textbf{Metaphor} & \textbf{AUC} \\
\midrule
16 & -0.020 & -0.015 & 0.034 & 0.8895\\
24 & -0.014 & -0.009 & 0.080 & 0.8375\\
32 & -0.004 & 0.020 & 0.080 & 0.9496\\
37 & 0.028 & 0.046 & 0.144 & 0.9296\\
41 & 0.017 & 0.029 & 0.125 & 0.9001\\
\bottomrule
\end{tabular}
\caption{
Mean $\Delta_{\mathrm{align}}$ across evaluated layers. 
The \textsc{Metaphor} set produces consistently larger representation-level shifts towards procedural behavior prototypes than \textsc{Benign} or \textsc{Neutral Distractor}. We use $\Delta_{\mathrm{align}}$ to distinguish \textsc{Metaphor} from the control sets. Higher AUC indicates stronger separability between metaphorical skills and the 
control sets.
}
\label{tab:proc-transfer-layers}
\end{table}

Table~\ref{tab:proc-transfer-layer37} reports the group-level alignment results at layer 37.The \textsc{Metaphor} set produces a substantially larger $\Delta_{\mathrm{align}}$ than both control sets. At this layer, \textsc{Metaphor} has a mean alignment score of $0.144$, compared with $0.028$ for \textsc{Benign} and $0.046$ for \textsc{Neutral Distractor} sets. This indicates that adding a metaphorical skill moves the model's representation alignment with the corresponding lower-efficiency procedural behavior prototype more strongly than adding benign or neutral background-domain guidance.

The difference is statistically large. At layer 37, \textsc{Metaphor} exceeds \textsc{Benign} by $0.116$ in mean $\Delta_{\mathrm{align}}$ (Cohen's $d=2.24$, Mann--Whitney $p=1.28\times10^{-26}$), and exceeds \textsc{Neutral Distractor} by $0.098$ ($d=1.88$, $p=2.76\times10^{-20}$). \textsc{Neutral Distractor} also shows a 
significant shift relative to \textsc{Benign} ($0.046$ vs. $0.028$), suggesting that background-domain framing can introduce weak procedural activation. However, this shift is much smaller than the shift induced by metaphorical skills, indicating that procedural metaphors have stronger impact on LLM's latent space.

\begin{table}[t]
\centering
\small
\begin{tabular}{lcccc}
\toprule
\textbf{Source set} & \textbf{N} & $\mathbf{A_{\mathrm{target}}}$ & $\mathbf{A_{\mathrm{joint}}}$ & $\mathbf{\Delta_{\mathrm{align}}}$ \\
\midrule
Benign & 90 & 0.321 & 0.349 & 0.028 \\
Neutral distractor & 90 & 0.321 & 0.367 & 0.046 \\
Metaphor & 90 & 0.307 & 0.451 & 0.144 \\
\bottomrule
\end{tabular}
\caption{
Representation-level behavior alignment at layer 37.
$A_{\mathrm{target}}=\cos(h_Q,h_B)$ is the similarity between the task-only representation and the behavior prototype, while $A_{\mathrm{joint}}=\cos(h_{S,Q},h_B)$ is the similarity after adding the skill instruction.
$\Delta_{\mathrm{align}}=A_{\mathrm{joint}}-A_{\mathrm{target}}$ measures how much the skill moves the representation towards the assigned procedural behavior prototype.
}
\label{tab:proc-transfer-layer37}
\end{table}

\paragraph{Layer-wise separability of metaphorical steering.}
We further evaluate whether the behavior alignment score can be used to distinguish metaphorical skills with negative impact from controls at the example level. For each evaluated layer, we use $\Delta_{\mathrm{align}}$ as a scalar score and compute the AUC for separating metaphorical skills from the combined \textbf{benign} and \textsc{Neutral Distractor} control sets. As shown in Table~\ref{tab:proc-transfer-layers}, $\Delta_{\mathrm{align}}$ separates metaphorical skills from controls across all evaluated layers, with AUC values ranging from $0.8375$ to $0.9496$. The strongest separability appears at layer 32, where AUC reaches $0.9496$. 

\begin{figure}[h!]
    \centering
    \includegraphics[width=\linewidth]{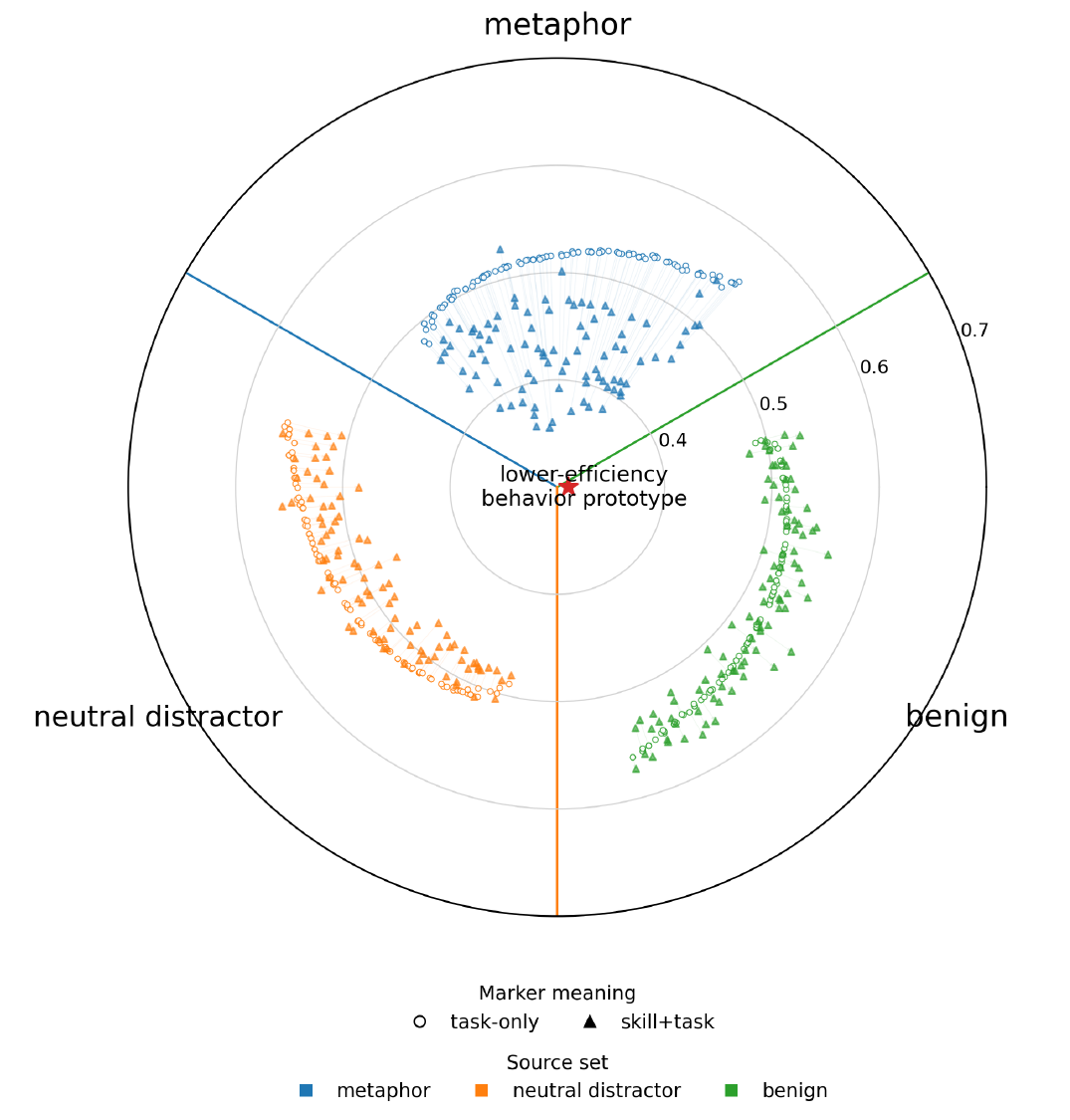}
     \caption{Prototype-centered radial shift plot (layer 32, zoomed to 0.7).  
     Each point pair corresponds to one skill-task example: open circles denote task-only representations, and filled triangles denote skill-conditioned representations. Arrows show the per-sample shift after adding the skill, with smaller radius indicating stronger alignment with the lower-efficiency behavior prototype.
     }
\label{fig:prototype_radial_plot}
\end{figure}

Figure \ref{fig:prototype_radial_plot} visualizes the prototype-centered radial shift at layer 32 for all 270 skill-task samples. For each example, the open circle denotes the task-only representation, while the filled triangle denotes the representation after adding the skill; arrows connect the two and indicate the direction of the representation shift. Smaller radius corresponds to stronger alignment with the lower-efficiency behavior prototype. The metaphor examples show a clear systematic inward shift after skill conditioning, indicating stronger alignment with the lower-efficiency prototype. In contrast, the neutral distractor examples exhibit a weaker shift, and the benign examples show little systematic movement towards the prototype. This suggests that metaphorical skills induce a distinctive representation/latent level movement towards lower-efficiency procedural behavior beyond what is observed for natural but non-steering controls.

These results indicate that the procedure-transfer signal in latent space is not only visible in group-level averages, but also provides strong example-level discrimination between negative metaphorical steering skills and non-steering controls. We further conduct representational similarity analysis to compare the internal representations across conditions, details see Appendix \ref{Appendix:RSA}.

\section{Defense Against Metaphorical Algorithmic Steering}

We formulate the defense as an instruction-level safety gate before code generation. Given a programming task $q$ and a skill instruction $s$, the detector predicts both the form of the skill and 
whether it is likely to induce the transfer of the target procedural behavior to the programming task, thus steering the model towards lower-efficiency algorithmic strategies.

\subsection{Experiment Setup}

We use the same \textsc{Metaphor}, \textsc{Benign}, and \textsc{Neutral Distractor} skill sets as in the representation-level analysis. In addition, we include a \textsc{Literal} skill set containing direct steering skills for descriptive comparison. Each skill set consists of 90 skills paired with the same 90 programming questions. The \textsc{Benign} set serves as a standard negative control, while the \textsc{Neutral Distractor} set provides a more challenging control by introducing natural background-domain framing without encouraging lower-efficiency algorithmic strategies.

We use Qwen-Coder-Next 80B as the detector.
For each task-skill pair $(q,s)$, the detector takes the skill text and the associated programming task as input. It is not given the full skill set, the skill-set category, record ID, file path, expected label, or any other metadata. All predictions are made solely from the input text. A condensed prompt provided in figure \ref{fig:detection-prompt}. 

For each $(q,s)$, we evaluate two dimensions of the defense gate. The first is \emph{instruction-form detection}, where the detector classifies each skill as \textsc{Metaphor}, \textsc{Literal}, \textsc{Fully-Benign}, or \textsc{Other}. This evaluates whether metaphorical steering can be recognized as a distinct instruction form. The second is \emph{low-efficiency impact prediction}, where the detector predicts whether the skill is likely to steer the code-generation model towards a lower-efficiency algorithmic strategy for the given programming task. For this evaluation, \textsc{Metaphor} skills are treated as positive examples, while the \textsc{Benign} and \textsc{Neutral Distractor} skill sets are treated as negative examples. The \textsc{Literal} set is excluded from the primary impact evaluation because its expected impact is mixed and therefore does not provide a clean ground-truth label.

In addition to these predictions, the detector produces a risk level together with mechanism-level explanations, indicating whether the skill encourages behaviors such as exhaustive enumeration, recomputation from scratch, repeated rescanning, anti-reuse, or unnecessary state expansion.

\begin{figure}[h!]
\centering
\begin{minipage}{0.98\columnwidth}
\hrule
\vspace{4pt}

{\ttfamily\scriptsize\RaggedRight
Classify the skill as one of:\par
\quad \path{cross_domain_metaphor_or_procedure_analogy},\par
\quad \path{literal_skill_steering}, 
\path{fully_benign}, or \path{ambiguous}.\par
\vspace{3pt}

Judge whether, for the given programming task, it is
likely to push a coder model toward code less efficient than a
natural straightforward solution. Mark impact only when the
skill creates causal pressure toward lower-efficiency behavior.\par
\vspace{3pt}

\textless examples\textgreater\par
\quad [Few-shot input and output demonstrations for cross-domain
metaphor, literal steering, and fully benign skills.]\par
\textless /examples\textgreater\par
\vspace{3pt}

Return category, impact label, confidence, evidence
spans, and a concise explanation.\par
}

\vspace{4pt}
\hrule
\end{minipage}
\caption{Condensed prompt used for skill-steering detection.}
\label{fig:detection-prompt}
\end{figure}

\begin{table}[h!]
\centering
\footnotesize
\setlength{\tabcolsep}{1mm}
\begin{tabular}{lcccc}
\toprule
\textbf{Class} & \textbf{Support} & \textbf{Precision} & \textbf{Recall} & \textbf{F1} \\
\midrule
Metaphor & 90 & 77.6\% & 100.0\% & 87.4\% \\
Literal & 90 & 95.3\% & 67.8\% & 79.2\% \\
Benign & 90 & 93.3\% & 93.3\% & 93.3\% \\
\midrule
Macro average & 270 & 88.7\% & 87.0\% & 86.6\% \\
Weighted average & 270 & 88.7\% & 87.0\% & 86.6\% \\
\bottomrule
\end{tabular}
\caption{Evaluation results of instruction-form detection.}
\label{tab:skill-steering-detection-metrics}
\end{table}

\begin{table}[h!]
\centering
\footnotesize
\setlength{\tabcolsep}{0.6mm}
\label{tab:low-efficiency-impact-metrics-v4}
\begin{tabular}{lcccc}
\toprule
\textbf{Class} & \textbf{Support} & \textbf{Precision} & \textbf{Recall} & \textbf{F1} \\
\midrule
Low-efficiency impact & 90 & 97.2\% & 77.8\% & 86.4\% \\
No low-efficiency impact & 180 & 89.9\% & 98.9\% & 94.2\% \\
\midrule
Macro average & 270 & 93.6\% & 88.3\% & 90.3\% \\
Weighted average & 270 & 92.3\% & 91.9\% & 91.6\% \\
\bottomrule
\end{tabular}
\caption{Evaluation for low-efficiency impact prediction. \textsc{No low-efficiency impact} class  contains \textsc{Benign} and \textsc{Neutral Distractor} groups. \textsc{Literal} group is excluded because its expected impact label is mixed.
}
\label{tab:low-efficiency-impact-metrics}
\end{table}

\subsection{Results}
Table~\ref{tab:skill-steering-detection-metrics} reports the results for instruction-form detection and Table~\ref{tab:low-efficiency-impact-metrics} reports the results of  algorithmic steering impact prediction. 
For instruction-form detection, the detector achieves a macro F1 of $86.6\%$ over 270 examples.
It identifies metaphorical steering skills with perfect recall ($100.0\%$), indicating that no metaphorical steering skills are missed. The detector also separates fully benign skills reliably, achieving an F1 score of $93.3\%$.

For algorithmic steering impact prediction, the detector achieves 
$90.3\%$ macro F1. 
It is especially precise when identifying skills with low-efficiency impact, achieving $97.2\%$ precision and an F1 score of $86.4\%$. 
The no-impact class also has high recall ($98.9\%$), showing that \textsc{Benign} and \textsc{Neutral Distractor} skills are rarely misclassified as having lower-efficiency impact.

These results suggest that the detector can reliably distinguish skills likely to induce lower-efficiency algorithmic steering from benign and neutral instructions, providing an effective instruction-level safety gate before code generation. More diagnostics on the skill-steering detector see Appendix \ref{Appendix:Additional_Diagnostics}.

\section{Conclusions}
In this work, we explored how  metaphorical expressions in skills can implicitly
transfer procedural patterns from source scenarios to code-generation tasks and steer models toward lower-efficiency algorithmic strategies. We further provided representation-level evidence of metaphorical algorithmic steering and developed an instruction-level method for detecting this vulnerability. Such subtle failures are triggered by the limitation of natural language and have a broad implication to using natural language to drive critical software design.


\bibliography{custom}


\section{Appendix}
\label{sec:appendix}

\subsection{Representational Similarity Analysis}\label{Appendix:RSA}
\begin{table}[h!]
\centering
\footnotesize
\setlength{\tabcolsep}{0.8mm}
\begin{tabular}{lccc}
\toprule
\textbf{Layer} & \textbf{Behavior RSA} & \textbf{Source-set RSA} & \textbf{Behavior $-$ source} \\
\midrule
16 & 0.537 & 0.293 & 0.243 \\
24 & 0.550 & 0.301 & 0.248 \\
32 & 0.537 & 0.314 & 0.223 \\
37 & 0.529 & 0.301 & 0.228 \\
41 & 0.517 & 0.299 & 0.218 \\
\bottomrule
\end{tabular}
\caption{
Representational similarity analysis across layers. Positive Behavior $-$ (minus) source values indicate that hidden state similarity is more aligned with procedural behavior labels than with source-set identity.
}
\label{tab:proc-transfer-rsa}
\end{table}

\paragraph{Representations cluster more by procedure than by source set.}
We further perform a representational similarity analysis (RSA) to test whether activation similarity is better explained by procedural behavior labels or by source-set identity. For each layer, we compute pairwise activation similarities between examples and compare them with two binary similarity matrices: one indicating whether two examples share the same procedural behavior label, and another indicating whether they come from the same source set. Across all evaluated layers, behavior similarity explains activation similarity more strongly than source-set identity. The behavior $-$(minus) source RSA gap ranges from $0.218$ to $0.248$, as shown in Table~\ref{tab:proc-transfer-rsa}. This suggests that the observed representation structure reflects procedure-level behavior rather than merely separating metaphorical, benign, and neutral inputs.

\paragraph{Summary.}
Together with the results in section \ref{Representation-Level-Analysis-results}, these results provide representation-level support for our procedural-transfer hypothesis. Metaphorical skills do not merely change the surface form of the prompt; they induce a measurable shift in hidden space towards lower-efficiency procedural behavior prototypes. The fact that neutral distractors produce only a much weaker shift suggests that the effect is not simply triggered by background-domain language. Instead, the strongest shifts occur when the skill contains a procedure that can be mapped from the source domain into the target programming task.

\subsection{Additional Diagnostics for the Skill-Steering Detector}\label{Appendix:Additional_Diagnostics}
\begin{table}[h!]
\centering
\footnotesize
\setlength{\tabcolsep}{0.8mm}
\begin{tabular}{lccc}
\toprule
\textbf{Expected $\backslash$ Detected} 
& \textbf{Metaphor} 
& \textbf{Literal} 
& \textbf{benign} \\
\midrule
Metaphor & 90 & 0 & 0 \\
Literal & 23 & 61 & 6 \\
benign & 3 & 3 & 84 \\
\bottomrule
\end{tabular}
\caption{Confusion matrix for instruction-form detection labels. Rows are expected labels and columns are detected labels.}
\label{tab:skill-steering-detection-confusion}
\end{table}

\begin{table}[h!]
\centering
\small
\begin{tabular}{lcc}
\toprule
\textbf{Expected impact $\backslash$ Predicted impact} & \textbf{true} & \textbf{false} \\
\midrule
true: metaphor & 70 & 20 \\
false: benign + neutral distractor & 2 & 178 \\
\bottomrule
\end{tabular}
\caption{Confusion matrix for low-efficiency impact prediction over source sets with known expected impact labels.}
\label{tab:skill-steering-impact-confusion}
\end{table}
The main text reports the two primary detector evaluations: instruction-form
detection and lower-efficiency impact prediction. Here we include the supporting
diagnostics behind those results.

\begin{table*}[ht]
\centering
\small
\begin{tabular}{lcccc}
\toprule
\textbf{Evaluation split} & \textbf{Precision} & \textbf{Recall} & \textbf{F1} & \textbf{Accuracy} \\
\midrule
metaphor vs benign & 100.0\% & 77.8\% & 87.5\% & 88.9\% \\
metaphor vs neutral distractor & 97.2\% & 77.8\% & 86.4\% & 87.8\% \\
metaphor vs benign + neutral distractor & 97.2\% & 77.8\% & 86.4\% & 91.9\% \\
\bottomrule
\end{tabular}
\caption{Low-efficiency impact prediction against source set expected impact labels. Literal baseline is excluded from metric computation because its expected impact label is mixed/unknown.}
\label{tab:skill-steering-impact-metrics}
\end{table*}

\begin{table*}[ht]
\centering
\small
\begin{tabular}{lccl}
\toprule
\textbf{Record id} & \textbf{Detected category} & \textbf{Risk level} & \textbf{Main explanation} \\
\midrule
311 & metaphor / procedure & medium & Interpreted ``repeated closing sweeps'' as full rescanning pressure. \\
376 & metaphor / procedure & medium & Interpreted record-keeping wording as discouraging compact summaries. \\
\bottomrule
\end{tabular}
\caption{False positives on the neutral distractor hard-negative set.}
\label{tab:skill-steering-neutral-fp}
\end{table*}

Table \ref{tab:skill-steering-detection-confusion} shows the confusion matrix for instruction-form detection. This result
is mainly a sanity check. It verifies that the detector can recognize the intended
form of the skills. The metaphorical skills are consistently identified as
cross-domain metaphor, and most benign controls are recognized as
fully benign. The literal baseline is less uniform, which is not surprising. Because literal baseline is a
comparison set for literal steering rather than a clean class for lower-efficiency impact.

Figures \ref{tab:skill-steering-impact-confusion} and \ref{tab:skill-steering-impact-metrics} focus on the impact prediction task. This output is more relevant to safety. Figure \ref{tab:skill-steering-impact-metrics} reports the split-level precision, recall, and
F1 scores, while Figure \ref{tab:skill-steering-impact-confusion} gives the corresponding confusion matrix for the main
impact split. The negative class here includes both ordinary benign controls and
neutral distractors. This is important because the neutral distractors are harder
negatives. They use background-domain or metaphorical analogy like phrasing, but they are
not designed to push the model toward a lower-efficiency solution. The detector
makes only two false-positive impact predictions among these negative examples,
suggesting that it is not simply reacting to metaphorical framing or procedural/analogical
language.

Figure \ref{tab:skill-steering-neutral-fp} inspects those two neutral-distractor false positives. In both cases,
the mistake is understandable, that the detector reads procedural phrasing more
strongly than intended. In one example, repeated ``closing sweeps'' is interpreted
as pressure toward full rescanning. In the other, record-keeping language is
interpreted as discouraging compact summaries. These cases point to detector's main
remaining failure mode that the metaphorical/procedural language can
occasionally look like efficiency-relevant steering, even when the skill is not
intended to change the algorithmic behavior.

\end{document}